\documentclass[sigconf]{acmart}
\usepackage[utf8]{inputenc}

\renewcommand\footnotetextcopyrightpermission[1]{} 
\setcopyright{none}
\acmDOI{}
\acmISBN{}

\acmConference[LND4IR]{Learning from Limited or Noisy Data for Information Retrieval}{12 July 2018}{Ann Arbor, MI, USA}
\acmYear{2018}
\copyrightyear{2018}
\acmPrice{15.00}

\begin{document}
\title{Explainable Agreement through Simulation for \\ Tasks with Subjective Labels}
\author{John Foley}
\affiliation{University of Massachusetts Amherst}
\email{jfoley@cs.umass.edu}

\begin{abstract}
    The field of information retrieval often works with limited and noisy data in an attempt to classify documents into subjective categories, e.g., relevance, sentiment and controversy. We typically quantify a notion of agreement to understand the difficulty of the labeling task, but when we present final results, we do so using measures that are unaware of agreement or the inherent subjectivity of the task.
   
    We propose using user simulation to understand the effect size of this noisy agreement data. By simulating truth and predictions, we can understand the maximum scores a dataset can support: for if a classifier is doing better than a reasonable model of a human, we cannot conclude that it is actually better, but that it may be learning noise present in the dataset.
    
    We present a brief case study on controversy detection that concludes that a commonly-used dataset has been exhausted: in order to advance the state-of-the-art, more data must be gathered at the current level of label agreement in order to distinguish between techniques with confidence.
\end{abstract}

\maketitle

\section{Introduction}

Although we often work with user-labeled data, and we can quantify inter-annotator agreement -- we often do not know what these agreement scores mean in terms of our target measures, like mean average precision (mAP), or F1. What does an agreement of 70\% allow for us to understand on this dataset? What level of mAP is believable? What is the best a system can do here?

Instead of an agreement probability or a statistical measure of agreement, what we really want is to be able to quantify the effect our level of agreement is having on our ability to evaluate systems. Therefore, what we really need is a technique for exploring what our agreement means in any arbitrary evaluation measure.

While existing agreement measures give us a sense of how difficult the task was for annotators, it is hard to quantify what that means for measures, particularly those involving ranking. Maybe annotators agree on the most important instances, or maybe they disagree on the most critical instances -- the same agreement score may lead to very different reliabilities in results.

Controversy is a problem that has attracted a lot of attention in recent years~\cite{dori2013detecting,jang2016probabilistic,jang2016improving,jang2017modeling,zielinski2018computing}. Much like relevance, sentiment, and other labels of interest, it is both somewhat subjective (noisy) and expensive to collect (limited). In this study, we will look at the dataset of 343 pages collected by Dori-Hacohen and Allan~\cite{dori2013detecting} and used in further studies~\cite{jang2016probabilistic,jang2016improving}. 

We find that the language-modeling approaches introduced by Jang et al.~\cite{jang2016probabilistic} effectively ``max-out'' this dataset, as their language modeling classifier achieves statistically indistinguishable performance from our human-model simulations. This means that given the limited dataset size and the inherent disagreement between annotators on which documents are controversial -- there are no AUC scores higher than currently published results that we should believe without collection of additional labels.

In this paper, we introduce a simulation technique that will allow this analysis to be performed on any dataset with a set of annotator labels for any arbitrary measure or metric. 

\section{Related Work}

The effect of the subjectivity and difficulty of relevance on IR evaluation has long been studied~\cite{bermingham2009study,carterette2010effect,webber2013assessor,voorhees2000variations,buckley2004retrieval,sanderson2005information}. While these studies look at the robustness of measures in the face of this subjectivity and noise, they do not quantify how well a system can do in comparison to humans -- probably because IR systems rarely retrieve otherwise perfect rankings.

As agreement measures can be used to evaluate classification tasks directly, studies connecting the two are often looking at the suitability of an agreement score for classifier evaluation, e.g,~\cite{ben2008relationship}.

Simulating users of IR or ML systems is also not a new contribution (e.g., \cite{tague1981simulation}) and recently work has begun to accelerate in this direction~\cite{maxwell2016agents}. However, we are unaware of work that simulates users in order to understand the limitations of agreement for a dataset.

\section{Truth Simulation Models}

In this section, we introduce a number of models for deriving truth from a set of labels for a document. 

Given a document $D$ which has a set of labels $L = \{l_1, l_2, \ldots l_{|L|}\}$, with each label $l_i$ being provided by a different annotator, most studies choose a simple heuristic function $f(L)$ that generates a single label from the set.

Our models are applicable to both binary and multiclass judgments, provided that functions $f(L)$ map from a set of labels to a valid label. Since we look at controversy, we focus on ordinal labels, and we can use fractional labels as predictions, but not truth.

\subsection{Average and Max Models}

In prior work~\cite{dori2013detecting,jang2016probabilistic,jang2016improving}, the assignment given to a document is the average of its labels.
Another appropriate model for controversy we consider in addition to the average modeling done here is a maximum model: i.e., a document is controversial if any annotator considered it controversial -- a policy aimed at maximizing recall. 

\subsection{Agreement-Flip Model}
Here we let $p$ be the probability of agreement calculated across the dataset. We could argue that with probability $1-p$, a label will be disputed and therefore is possibly incorrect with this probability. This is a fairly simple model of agreement, and the one that is represented by presenting agreement ratio in papers.

\subsection{Label Sampling Model}
A better model takes document-level confusion into account: if a document garners a variety of labels, we consider these observations of the underlying distribution for that document. Here, our $f$ samples a label at random from a document.

\subsection{Label Conflation Model}

In a world where there are multi-value relevance labels, and you have many documents with only a single annotator e.g., Excellent, Good, Fair, Bad, we may wish to have a simulation that can generalize to these cases in a more accurate manner.

Our label conflation model first learns the probabilities of mistaking labels for each other. We would expect disagreement between highly-relevant and relevant documents, for instance, but less disagreement between highly-relevant and non-relevant documents. However, this model is data driven, so it will reflect the actual behavior of users.
As a concrete example, the model learned for labels in the Dori-Hacohen and Allan dataset is presented in Table~\ref{tab:conflation}. 

Given any truth label, we then sample a new value based on how humans often disagree with that particular truth value.

\begin{table}[ht]
    \caption{Label Conflation Model for Controversy Web Pages~\cite{dori2013detecting}. {\rm The label with most agreement is the ``Clearly Non-Controversial'' label.}}
    \label{tab:conflation}
    \centering
    \begin{tabular}{lr|rrrr}
        & & \multicolumn{4}{c}{Conflated With} \\
        Text  & \# & 2 & 1 & 0 & -1  \\ \hline
        Very Controversial & 2  & 237 & 83 & 23 & 48 \\
        Controversial & 1  & 83 & 182 & 27 & 53 \\
        Possibly Non-Controversial & 0  & 23 & 27 & 133 & 92 \\
        Clearly Non-Controversial & -1 & 48 & 53 & 92 & 594 \\
    \end{tabular}
\end{table}
\section{Results}
Given our set of models that each reasonably approximate human labeling disagreement on this ambiguous task, we can now run a simulation to understand what the expected performance (under any measure) should be for our humans under these models. For each setting, we run $N=10000$ simulations.

\begin{table}[ht]
    \caption{Simulated AUC scores for controversy detection. {\rm When predictions and truth are generated by the given models, we obtain a the following sampling of AUC scores.}}
    \label{tab:simulationAUCs}
    \centering
    \begin{tabular}{lcc|rrr}
        & & & \multicolumn{3}{c}{Percentile AUC} \\
        \# & System Model & Truth Model & 5th & 50th & 95th\\ \hline
        1 & Sample & Average & 0.862 & 0.890 & 0.917 \\
        2 & Sample & Max & 0.846 & 0.874 & 0.900 \\
        3 & Sample & Sample & 0.818 & 0.852 & 0.884 \\
        4 & Conflate(Truth) & Sample & 0.794 & 0.836 & 0.875 \\
        5 & Conflate(Sample) & Conflate(Sample) & 0.674 & 0.725 & 0.774 \\
        6 & Flip(p=0.643, Truth) & Average & 0.593 & 0.639 & 0.685 \\
    \end{tabular}
\end{table}

In prior work, the best AUC reported for this task is 0.856~\cite{jang2016probabilistic}, and the AUC reported for the original work is 0.743~\cite{dori2013detecting}. We present six pairings of our truth simulation models in Table~\ref{tab:simulationAUCs}. 
We have ordered our simulations from optimistic (label sampling system, average truth \#1) to pessimistic (traditional agreement probabilities \#6). 
This suggests to us that we can believe in the improvement presented from 0.743-0.856, but that we should be skeptical of any further improvements shown on this dataset, as even our optimistic models suggest that we are doing as well as a human can do given the ambiguity of the task.

\section{Conclusions}
In this work, we have briefly presented a number of strategies for investigating the agreement of users on labeling tasks. Given a vector of document labels assigned by different people at each document, we can model the difficulty of particular instances and particular labels. Further work is needed to understand the best simulation models for given tasks, but exploring a variety of reasonable models allows us to come to a reasonable conclusion that the discriminative power of an existing controversy detection dataset has been used up in terms of a robust classification metric: AUC. We therefore propose that future work on classifying or ranking using subjective labels consider simulation as an explainable alternative to opaque agreement scores.

\section*{Acknowledgements}

This work was supported in part by the Center for Intelligent  Information  Retrieval. 

\bibliographystyle{ACM-Reference-Format}
\bibliography{cites}


\begin{thebibliography}{00}


\ifx \showCODEN    \undefined \def \showCODEN     #1{\unskip}     \fi
\ifx \showDOI      \undefined \def \showDOI       #1{#1}\fi
\ifx \showISBNx    \undefined \def \showISBNx     #1{\unskip}     \fi
\ifx \showISBNxiii \undefined \def \showISBNxiii  #1{\unskip}     \fi
\ifx \showISSN     \undefined \def \showISSN      #1{\unskip}     \fi
\ifx \showLCCN     \undefined \def \showLCCN      #1{\unskip}     \fi
\ifx \shownote     \undefined \def \shownote      #1{#1}          \fi
\ifx \showarticletitle \undefined \def \showarticletitle #1{#1}   \fi
\ifx \showURL      \undefined \def \showURL       {\relax}        \fi
\providecommand\bibfield[2]{#2}
\providecommand\bibinfo[2]{#2}
\providecommand\natexlab[1]{#1}
\providecommand\showeprint[2][]{arXiv:#2}

\bibitem[\protect\citeauthoryear{Ben-David}{Ben-David}{2008}]%
        {ben2008relationship}
\bibfield{author}{\bibinfo{person}{Arie Ben-David}.}
  \bibinfo{year}{2008}\natexlab{}.
\newblock \showarticletitle{About the relationship between ROC curves and
  Cohen's kappa}.
\newblock \bibinfo{journal}{{\em Engineering Applications of Artificial
  Intelligence\/}} \bibinfo{volume}{21}, \bibinfo{number}{6}
  (\bibinfo{year}{2008}), \bibinfo{pages}{874--882}.
\newblock


\bibitem[\protect\citeauthoryear{Bermingham and Smeaton}{Bermingham and
  Smeaton}{2009}]%
        {bermingham2009study}
\bibfield{author}{\bibinfo{person}{Adam Bermingham} {and}
  \bibinfo{person}{Alan~F. Smeaton}.} \bibinfo{year}{2009}\natexlab{}.
\newblock \showarticletitle{A Study of Inter-annotator Agreement for Opinion
  Retrieval}. In \bibinfo{booktitle}{{\em SIGIR}}. \bibinfo{pages}{784--785}.
\newblock


\bibitem[\protect\citeauthoryear{Buckley and Voorhees}{Buckley and
  Voorhees}{2004}]%
        {buckley2004retrieval}
\bibfield{author}{\bibinfo{person}{Chris Buckley} {and}
  \bibinfo{person}{Ellen~M Voorhees}.} \bibinfo{year}{2004}\natexlab{}.
\newblock \showarticletitle{Retrieval evaluation with incomplete information}.
  In \bibinfo{booktitle}{{\em SIGIR}}. \bibinfo{pages}{25--32}.
\newblock


\bibitem[\protect\citeauthoryear{Carterette and Soboroff}{Carterette and
  Soboroff}{2010}]%
        {carterette2010effect}
\bibfield{author}{\bibinfo{person}{Ben Carterette} {and} \bibinfo{person}{Ian
  Soboroff}.} \bibinfo{year}{2010}\natexlab{}.
\newblock \showarticletitle{The effect of assessor error on IR system
  evaluation}. In \bibinfo{booktitle}{{\em SIGIR}}. \bibinfo{pages}{539--546}.
\newblock


\bibitem[\protect\citeauthoryear{Dori-Hacohen and Allan}{Dori-Hacohen and
  Allan}{2013}]%
        {dori2013detecting}
\bibfield{author}{\bibinfo{person}{Shiri Dori-Hacohen} {and}
  \bibinfo{person}{James Allan}.} \bibinfo{year}{2013}\natexlab{}.
\newblock \showarticletitle{Detecting controversy on the web}. In
  \bibinfo{booktitle}{{\em CIKM}}. \bibinfo{pages}{1845--1848}.
\newblock


\bibitem[\protect\citeauthoryear{Jang and Allan}{Jang and Allan}{2016}]%
        {jang2016improving}
\bibfield{author}{\bibinfo{person}{Myungha Jang} {and} \bibinfo{person}{James
  Allan}.} \bibinfo{year}{2016}\natexlab{}.
\newblock \showarticletitle{Improving automated controversy detection on the
  web}. In \bibinfo{booktitle}{{\em SIGIR}}. ACM, \bibinfo{pages}{865--868}.
\newblock


\bibitem[\protect\citeauthoryear{Jang, Dori-Hacohen, and Allan}{Jang
  et~al\mbox{.}}{2017}]%
        {jang2017modeling}
\bibfield{author}{\bibinfo{person}{Myungha Jang}, \bibinfo{person}{Shiri
  Dori-Hacohen}, {and} \bibinfo{person}{James Allan}.}
  \bibinfo{year}{2017}\natexlab{}.
\newblock \showarticletitle{Modeling Controversy within Populations}. In
  \bibinfo{booktitle}{{\em ICTIR}}. ACM, \bibinfo{pages}{141--149}.
\newblock


\bibitem[\protect\citeauthoryear{Jang, Foley, Dori-Hacohen, and Allan}{Jang
  et~al\mbox{.}}{2016}]%
        {jang2016probabilistic}
\bibfield{author}{\bibinfo{person}{Myungha Jang}, \bibinfo{person}{John Foley},
  \bibinfo{person}{Shiri Dori-Hacohen}, {and} \bibinfo{person}{James Allan}.}
  \bibinfo{year}{2016}\natexlab{}.
\newblock \showarticletitle{Probabilistic approaches to controversy detection}.
  In \bibinfo{booktitle}{{\em CIKM}}. \bibinfo{pages}{2069--2072}.
\newblock


\bibitem[\protect\citeauthoryear{Maxwell and Azzopardi}{Maxwell and
  Azzopardi}{2016}]%
        {maxwell2016agents}
\bibfield{author}{\bibinfo{person}{David Maxwell} {and} \bibinfo{person}{Leif
  Azzopardi}.} \bibinfo{year}{2016}\natexlab{}.
\newblock \showarticletitle{Agents, simulated users and humans: An analysis of
  performance and behaviour}. In \bibinfo{booktitle}{{\em CIKM}}.
  \bibinfo{pages}{731--740}.
\newblock


\bibitem[\protect\citeauthoryear{Sanderson and Zobel}{Sanderson and
  Zobel}{2005}]%
        {sanderson2005information}
\bibfield{author}{\bibinfo{person}{Mark Sanderson} {and}
  \bibinfo{person}{Justin Zobel}.} \bibinfo{year}{2005}\natexlab{}.
\newblock \showarticletitle{Information retrieval system evaluation: effort,
  sensitivity, and reliability}. In \bibinfo{booktitle}{{\em SIGIR}}. ACM,
  \bibinfo{pages}{162--169}.
\newblock


\bibitem[\protect\citeauthoryear{Tague and Nelson}{Tague and Nelson}{1981}]%
        {tague1981simulation}
\bibfield{author}{\bibinfo{person}{Jean~M Tague} {and}
  \bibinfo{person}{Michael~J Nelson}.} \bibinfo{year}{1981}\natexlab{}.
\newblock \showarticletitle{Simulation of user judgments in bibliographic
  retrieval systems}. In \bibinfo{booktitle}{{\em ACM SIGIR Forum}},
  Vol.~\bibinfo{volume}{16}. ACM, \bibinfo{pages}{66--71}.
\newblock


\bibitem[\protect\citeauthoryear{Voorhees}{Voorhees}{2000}]%
        {voorhees2000variations}
\bibfield{author}{\bibinfo{person}{Ellen~M Voorhees}.}
  \bibinfo{year}{2000}\natexlab{}.
\newblock \showarticletitle{Variations in relevance judgments and the
  measurement of retrieval effectiveness}.
\newblock \bibinfo{journal}{{\em Information processing \& management\/}}
  \bibinfo{volume}{36}, \bibinfo{number}{5} (\bibinfo{year}{2000}),
  \bibinfo{pages}{697--716}.
\newblock


\bibitem[\protect\citeauthoryear{Webber and Pickens}{Webber and
  Pickens}{2013}]%
        {webber2013assessor}
\bibfield{author}{\bibinfo{person}{William Webber} {and}
  \bibinfo{person}{Jeremy Pickens}.} \bibinfo{year}{2013}\natexlab{}.
\newblock \showarticletitle{Assessor Disagreement and Text Classifier
  Accuracy}. In \bibinfo{booktitle}{{\em SIGIR}}. \bibinfo{pages}{929--932}.
\newblock


\bibitem[\protect\citeauthoryear{Zielinski, Nielek, Wierzbicki, and
  Jatowt}{Zielinski et~al\mbox{.}}{2018}]%
        {zielinski2018computing}
\bibfield{author}{\bibinfo{person}{Kazimierz Zielinski},
  \bibinfo{person}{Radoslaw Nielek}, \bibinfo{person}{Adam Wierzbicki}, {and}
  \bibinfo{person}{Adam Jatowt}.} \bibinfo{year}{2018}\natexlab{}.
\newblock \showarticletitle{Computing controversy: Formal model and algorithms
  for detecting controversy on Wikipedia and in search queries}.
\newblock \bibinfo{journal}{{\em Information Processing \& Management\/}}
  \bibinfo{volume}{54}, \bibinfo{number}{1} (\bibinfo{year}{2018}),
  \bibinfo{pages}{14--36}.
\newblock


\end{thebibliography}

\end{document}